\documentclass[twocolumn,showpacs,showkeys,reprint,amsmath,amssymb,aps]{revtex4-1}

\usepackage{graphicx}
\usepackage{dcolumn}
\usepackage{bm}
\usepackage{tabularx}
\usepackage{amsmath}
\usepackage{xcolor}

\begin{document}

\title{Double-dome superconductivity under pressure in the V-based Kagome metals $A$V$_3$Sb$_5$ ($A$ = Rb and K)}

\author{C. C. Zhu,$^{1}$, X. F. Yang,$^{1,\sharp}$, W. Xia,$^{2,3}$ Q. W. Yin,$^4$ L. S. Wang,$^{1}$ C. C. Zhao,$^{1}$ D. Z. Dai,$^{1}$ C. P. Tu,$^{1}$ B. Q. Song,$^{1}$ Z. C. Tao,$^2$ Z. J. Tu,$^4$ C. S. Gong,$^4$ H. C. Lei$^{4,\dag}$ Y. F. Guo,$^{2,\ddag}$ and S. Y. Li$^{1,5,6*}$}

\affiliation
 {$^1$State Key Laboratory of Surface Physics, Department of Physics, Fudan University, Shanghai 200438, China\\
 $^2$School of Physical Science and Technology, ShanghaiTech University, Shanghai 201210, China\\
 $^3$ShanghaiTech Laboratory for Topological Physics, Shanghai 201210, China\\
 $^4$Department of Physics and Beijing Key Laboratory of Opto-electronic Functional Materials and Micro-nano Devices, Renmin University of China, Beijing 100872, China\\
 $^5$Collaborative Innovation Center of Advanced Microstructures, Nanjing 210093, China\\
 $^6$Shanghai Research Center for Quantum Sciences, Shanghai 201315, China
}

\date{\today}

\begin{abstract}
We present high-pressure resistance measurements on the newly discovered V-based superconductors $A$V$_3$Sb$_5$ ($A$ = Rb and K), which have an ideal Kagome lattice of vanadium. Two superconducting domes under pressure are observed in both compounds, as previously observed in their sister compound CsV$_3$Sb$_5$. For RbV$_3$Sb$_5$, the $T_c$ increases from 0.93 K at ambient pressure to the maximum of 4.15 K at 0.38 GPa in the first dome. The second superconducting dome has the highest $T_c$ of 1.57 K at 28.8 GPa. KV$_3$Sb$_5$ displays a similar double-dome phase diagram, however, its two maximum $T_c$s are lower, and the $T_c$ drops faster in the second dome than RbV$_3$Sb$_5$. An integrated temperature-pressure phase diagram of $A$V$_3$Sb$_5$ ($A$ = Cs, Rb and K) is constructed, showing that the ionic radius of the intercalated alkali-metal atoms has a significant effect. Our work demonstrates that double-dome superconductivity under pressure is a common feature of these V-based Kagome metals.
\end{abstract}

\maketitle

The materials with geometrically frustrated Kagome lattice, depending on the degree of electron filling, host a wide range of intriguing physics, including quantum spin liquid, charge density wave (CDW), topological states, and superconductivity \cite{Spin-liquid,van Hove,Topological insulator,Dirac metal,Dirac fermions,Doped kagome}. Recently, a new family of V-based Kagome-lattice compounds $A$V$_3$Sb$_5$ ($A$ = Cs, Rb and K) was discovered \cite{Rb-K-Cs}. The CDW order manifests in magnetic susceptibility, resistivity, and specific heat at 94, 103, and 78 K for $A$ = Cs, Rb, and K, respectively \cite{CsV3Sb5-Z2,KVSb-Z2,RbVSb-SC,topological_charge_order_KVSb}. At low temperature, these compounds also show superconductivity with superconducting transition temperature (${\it T}_{\rm c}$) of 2.5, 0.92, and 0.93 K for $A$ = Cs, Rb, and K \cite{CsV3Sb5-Z2,KVSb-Z2,RbVSb-SC}. Interestingly, the scanning tunnelling microscope (STM) study demonstrates a topological charge order in KV$_3$Sb$_5$ \cite{topological_charge_order_KVSb}, which may cause the giant anomalous Hall effect \cite{anomalous Hall effect} and possibility unconventional superconductivity \cite{topological_charge_order_KVSb}. The angle-resolved photoemission spectroscopy (ARPES) and density-functional theory demonstrated the existence of ${\it Z}_{\rm 2}$ nontrivial topological band structure in CsV$_3$Sb$_5$ \cite {CsV3Sb5-Z2}.

For the superconducting gap structure in CsV$_3$Sb$_5$, while thermal conductivity measurements suggested nodal superconductivity \cite{SYLiCsVSb}, penetration depth and nuclear magnetic resonance (NMR) measurements claimed nodeless $s$-wave superconductivity \cite{Penetration,NMR}. The latest ultra-low temperature STM measurements shows both nodal and nodeless gaps \cite{STM-DLFeng} in this superconductor with multiple Fermi surfaces \cite{QuanOsci}. Theoretically, dominant $f$-wave triplet superconducting order, succeeded by $d$-wave singlet pairing for stronger coupling was found over a large range of coupling strength \cite{fwave}. In this context, the exact locations of the gap nodes in CsV$_3$Sb$_5$ still need to be identified.

Meanwhile, two superconducting domes were observed in CsV$_3$Sb$_5$ under pressure, together with the nodal superconductivity, further indicating unconventional superconductivity in CsV$_3$Sb$_5$ \cite{SYLiCsVSb}. The origin of the first dome in CsV$_3$Sb$_5$  is likely related to the CDW instability, due to the competition between CDW and superconductivity \cite{SYLiCsVSb,J.-G. Cheng}. The second superconducting dome was attributed to a pressure-induced Lifshitz transition and enhanced electron-phonon coupling in CsV$_3$Sb$_5$ \cite{Z. Yang,X. Chen}. To understanding this double-dome phase diagram, it will be interesting to study $A$V$_3$Sb$_5$ ($A$ = Rb and K) under pressure and compare with CsV$_3$Sb$_5$.

In this Letter, we performed high-pressure resistance measurements on $A$V$_3$Sb$_5$ ($A$ = Rb and K) up to 50 GPa. A clear double-dome superconductivity was revealed in the temperature-pressure phase diagram of both compounds, as observed in CsV$_3$Sb$_5$. The ionic radius of the intercalated alkali-metal atoms apparently affects the phase diagram of $A$V$_3$Sb$_5$. We discuss the possible pairing mechanism under the two superconducting domes.

\begin{figure}
\includegraphics[clip,width=8cm]{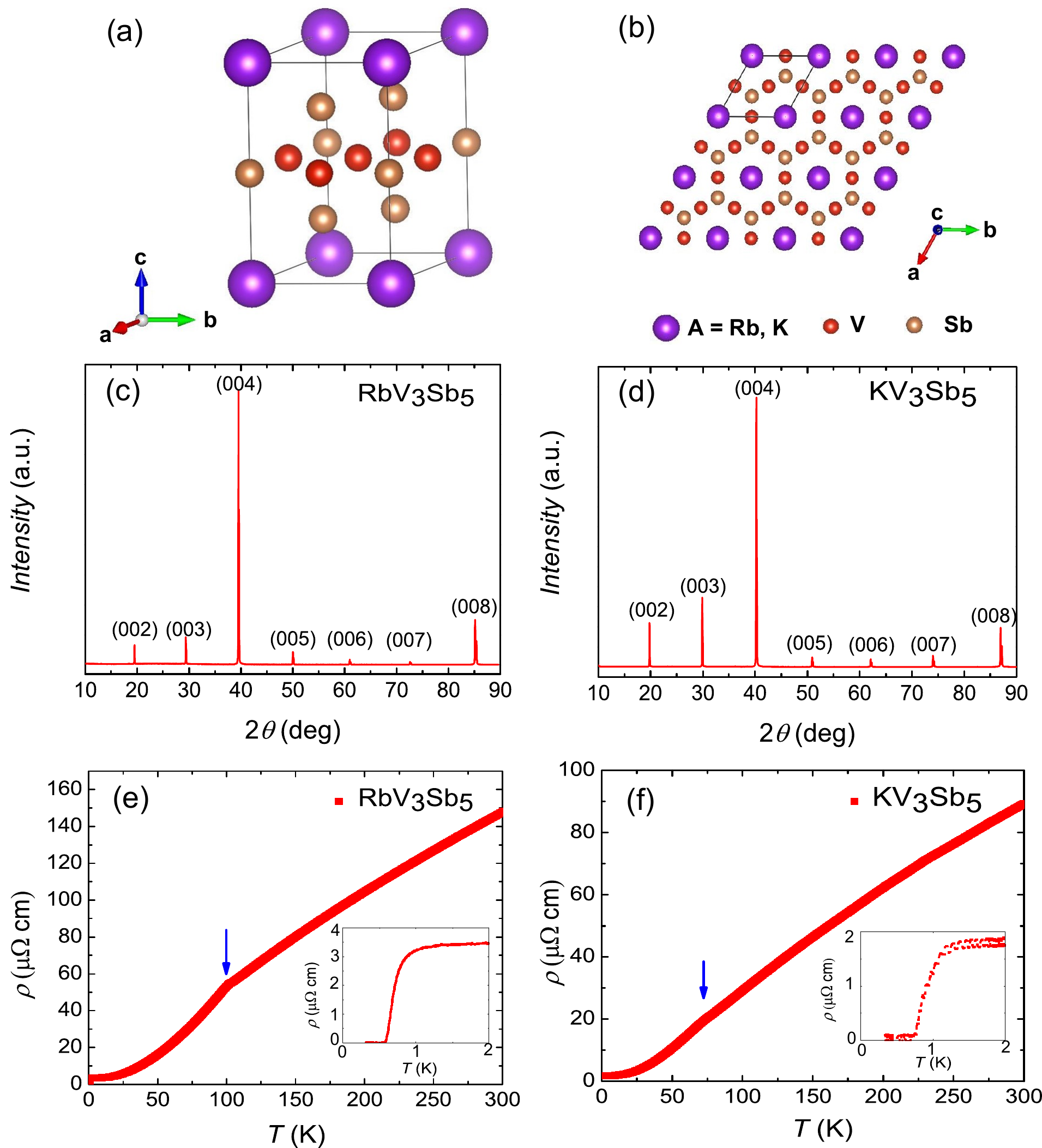}
\caption{(a) Crystal structure of $A$V$_3$Sb$_5$ ($A$ = Cs, Rb and K). The $A$, V, Sb atoms are presented as purple, orange and yellow balls, respectively. (b) Top view of the crystal structure, which shows the Kagome lattice of vanadium. (c) and (d) X-ray diffraction pattern for the largest surface of the $A$V$_3$Sb$_5$ ($A$ = Rb and K) single crystals. (e) and (f) Resistivity curves for RbV$_3$Sb$_5$ and KV$_3$Sb$_5$ single crystals respectively at ambient pressure. Blue arrows denote the CDW transitions. Insets of (e) and (f): superconducting transitions for $A$V$_3$Sb$_5$ ($A$ = Rb and K).}
\end{figure}

\begin{figure}
\includegraphics[clip,width=8.5cm]{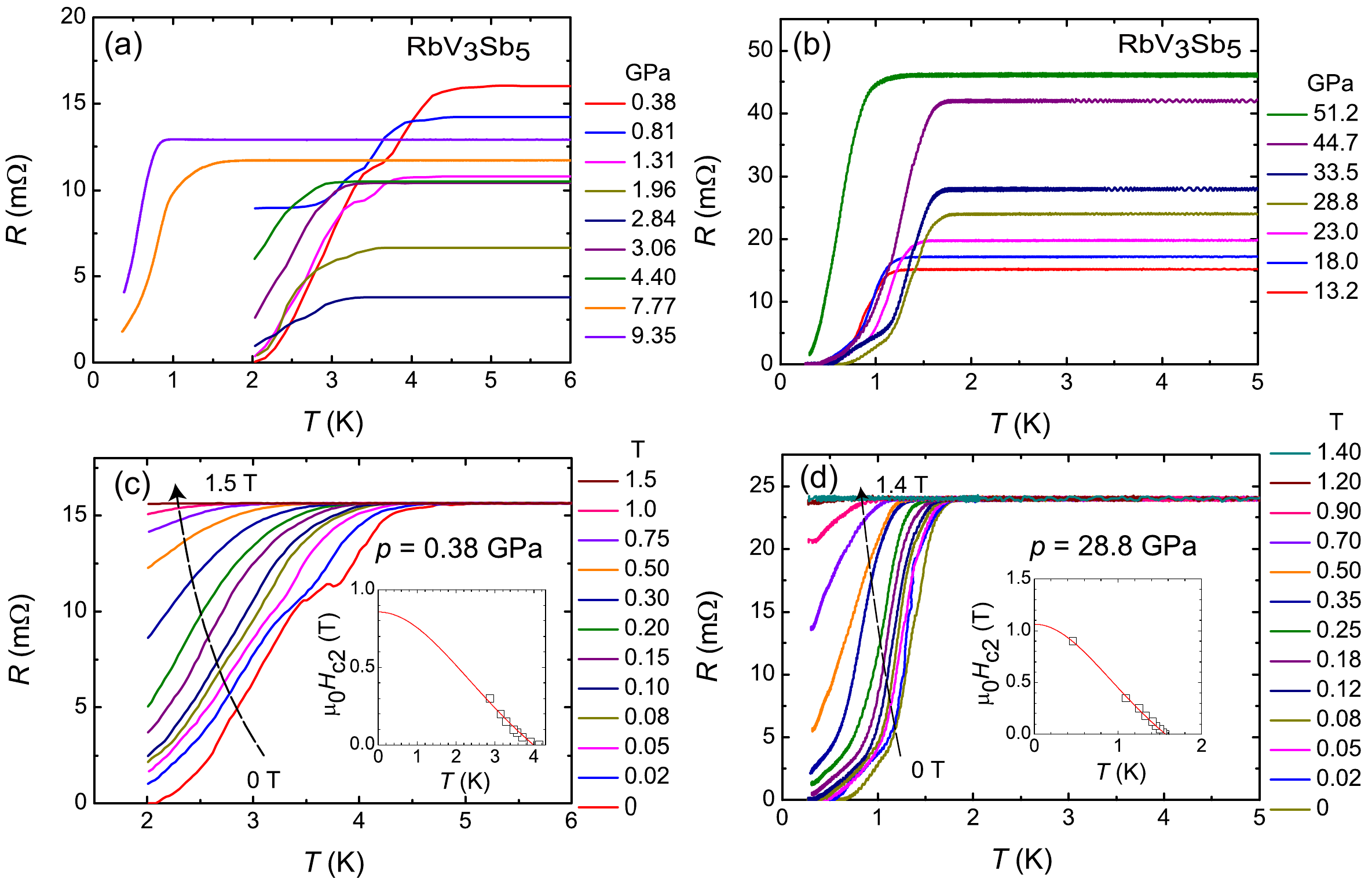}
\caption{(a) Temperature dependence of resistance for RbV$_3$Sb$_5$ under various pressures up to 9.35 GPa. (b) Temperature dependence of resistance for RbV$_3$Sb$_5$ under higher pressures up to 51.2 GPa. (c) and (d) Temperature dependence of resistance for RbV$_3$Sb$_5$ under different magnetic fields at 0.38 GPa and 28.8 GPa, respectively. Increasing the magnetic field gradually suppresses the superconducting transition. Inset: Temperature dependence of the upper field $\mu$$_0$${\it H}_{\rm c2}$. The superconducting transition temperature ${\it T}_{\rm c}$ is defined as the 10\% drop of the normal-state resistance (${\it T}_{\rm c}$$^{10\%}$). The red line shows the fitting of Ginzburg-Landau theory, $\mu$$_0$${\it H}_{\rm c2}$($T$)= $\mu$$_0$${\it H}_{\rm c2}$(0)(1-($T$/${\it T}_{\rm c}$)$^2$)/(1+($T$/${\it T}_{\rm c}$)$^2$). }
\end{figure}

Single crystal of $A$V$_3$Sb$_5$ ($A$ = Rb and K) were grown from K ingot (purity 99.8\%) or Rb (purity 99.75\%), V powder (purity 99.9\%), Sb powder (purity 99.999\%) through the self-flux method \cite{Rb-K-Cs}. For RbV$_3$Sb$_5$, the eutectic mixture of RbSb and Rb$_3$Sb$_7$ was mixed with VSb$_2$ to form a composition with 50 at.\% $A$$_x$Sb$_y$ ($A$ = Rb and K) and 50 at.\% VSb$_2$ approximately. The mixture was loaded into an alumina crucible which was sealed in a quartz ampoule under partial argon atmosphere followed with heating to 1273 K by taking 12 h and kept there for 24 h. Subsequently, the temperature was cooled down to 1173 K at 50 K/h and further to 923 K with a slow rate. Finally, the ampoule was taken out from the furnace and decanted with a centrifuge to separate RbV$_3$Sb$_5$ single crystal from the flux. Analogous reactions were performed for  KV$_3$Sb$_5$. The X-ray diffraction (XRD) was performed using powder X-ray diffractometer (D8 Advance, Bruker) with Cu $K$$\alpha$ radiation (wave length $\lambda$ = 1.5418 \AA).

High pressure was generated by a diamond anvil cell (DAC) made from Be-Cu alloy with two opposite diamond anvils. The diamonds with 300 $\mu$m culets were used for this experiment. Single crystals were crashed into powders and loaded inside of the DAC. The Be-Cu plate was used as gasket and cubic boron nitride served as insulating materials. The pressure inside of the DAC were scaled by monitoring the Ruby fluorescence at room temperature each time before and after the measurement. High pressure resistance of $A$V$_3$Sb$_5$ ($A$ = Rb and K) were carried out in a physical property measurement system (PPMS, Quantum Design) and a $^3$He cryostat with Van der Pauw method.

As depicted in Figs. 1(a) and 1(b), $A$V$_3$Sb$_5$ ($A$ = Cs, Rb and K) shows a layered structure with hexagonal symmetry ($P$6/$mmm$, No.191). V ions are coordinated by Sb atoms, forming slabs with ideal Kagome structure. The V-Sb layers are stacked along $c$ axis separated by alkali-metal layers. The lattice constant $c$ is 9.073(3) {\AA} and 8.943(1) {\AA} for Rb and K, respectively \cite{Rb-K-Cs}. Figures 1(c) and 1(d) show XRD patterns of $A$V$_3$Sb$_5$ single crystals, in which only ($00l$) Bragg peaks can be observed, indicating that the largest natural face of $A$V$_3$Sb$_5$ single crystals is the $ab$ plane. Typical resistivity curves of $A$V$_3$Sb$_5$ single crystals at ambient pressure are plotted in Figs. 1(e) and 1(f). The residual resistivity ratio $RRR$ = $\rho$(300 K)/$\rho$(1.5 K) is 42 and 51 for RbV$_3$Sb$_5$ and KV$_3$Sb$_5$, respectively, suggesting high quality of the samples. A kink is observed at 100 K for RbV$_3$Sb$_5$ and 75 K for KV$_3$Sb$_5$, corresponding to the CDW transition as reported previously \cite{KVSb-Z2,RbVSb-SC,topological_charge_order_KVSb}. The ${\it T}_{\rm c}$ defined at 10\% drop of normal-state resistance (${\it T}_{\rm c}$$^{10\%}$) is 0.93 and 1.24 K for RbV$_3$Sb$_5$ and KV$_3$Sb$_5$, respectively, as shown in the inset of Figs. 1(e) and 1(f), consistent with previous report \cite{KVSb-Z2,RbVSb-SC}.

Figure 2(a) presents the low-temperature resistance of RbV$_3$Sb$_5$ in a pressure range of 0.38\textendash 9.35 GPa. The ${\it T}_{\rm c}$$^{10\%}$ of RbV$_3$Sb$_5$ remarkably increases from 0.93 K at ambient pressure to 4.15 K at 0.38 GPa, then shifts to lower temperature with increasing pressure until reaches to 0.86 K at 9.35 GPa, which is similar to that of CsV$_3$Sb$_5$ \cite{SYLiCsVSb}. By applying a slightly higher pressure, zero-resistance behaviour re-emerges and ${\it T}_{\rm c}$$^{10\%}$ starts to increase with pressure to a maximum of 1.57 K at 28.8 GPa (Fig. 2(b)), showing a second superconducting dome.

To confirm that the resistance drop in RbV$_3$Sb$_5$ under pressures is superconducting transition, various magnetic fields are applied at 0.38 and 28.8 GPa (Figs. 2(c) and 2(d)). The resistance drop is monotonically suppressed with increasing magnetic field as expected and completely vanishes in 1.5 T at 0.38 GPa and 1.4 T at 28.8 GPa, demonstrating that the resistance drop under pressure is indeed attributed to superconducting transition. The upper critical field ($\mu$$_0$${\it H}_{\rm c2}$) versus $T$$_c$$^{10\%}$ of RbV$_3$Sb$_5$ at 0.38 and 28.8 GPa are summarized in the insets of Figs. 2(c) and 2(d). The data can be well fitted by the empirical Ginzburg-Landau (GL) formula $\mu$$_0$${\it H}_{\rm c2}$($T$)= $\mu$$_0$${\it H}_{\rm c2}$(0)(1-($T$/${\it T}_{\rm c}$)$^2$)/(1+($T$/${\it T}_{\rm c}$)$^2$). The $\mu$$_0$${\it H}_{\rm c2}$(0) are determined to be 0.856 and 1.063 T for 0.38 and 28.8 GPa, respectively. These values are much lower than the Pauli paramagnetic limit field $H$$_p$(0) = 1.84${\it T}_{\rm c}$ $\approx$ 7.6 and 2.89 T, indicating the absence of Pauli pair breaking.

In analogy with that of RbV$_3$Sb$_5$, the resistance of KV$_3$Sb$_5$ under pressures is also measured and plotted in Fig. 3. Figures 3(a) and 3(b) show the low-temperature resistance of KV$_3$Sb$_5$ in the pressure range of 0.32\textendash 3.68 GPa and 7.5\textendash 41.5 GPa, respectively. Initially, the ${\it T}_{\rm c}$$^{10\%}$ of KV$_3$Sb$_5$ increases sharply from 1.24 K at ambient pressure to 2.9 K at 0.32 GPa, which is, however, lower than those of RbV$_3$Sb$_5$ and CsV$_3$Sb$_5$ \cite{SYLiCsVSb}. With increasing pressure, the resistance drop gradually moves towards lower temperature until reaches the minimum of 0.58 K at 3.68 GPa, which is similar to previous report \cite{Pressure-KVSb}. Upon further compression, the ${\it T}_{\rm c}$$^{10\%}$ increases to 1.14 K around 20 GPa followed with slowly reduction and vanishes below 0.3 K at 41.5 GPa. Therefore, a second superconducting dome is also observed in KV$_3$Sb$_5$. The magnetic fields are applied to the sample at 0.32 and 14.7 GPa, as seen in Figs. 3(c) and 3(d). The resistance drops completely suppresses at 0.8 T for 0.32 GPa and 0.5 T for 14.7 GPa. Moreover, the $\mu$$_0$${\it H}_{\rm c2}$\textendash $T$ plots are well fitted with GL equation with the fit values of $\mu$$_0$${\it H}_{\rm c2}$(0) being 0.307 T for 0.32 GPa and 0.42 T for 14.7 GPa in insets of Figs. 3(c) and (d)), respectively.

\begin{figure}
\includegraphics[clip,width=8.5 cm]{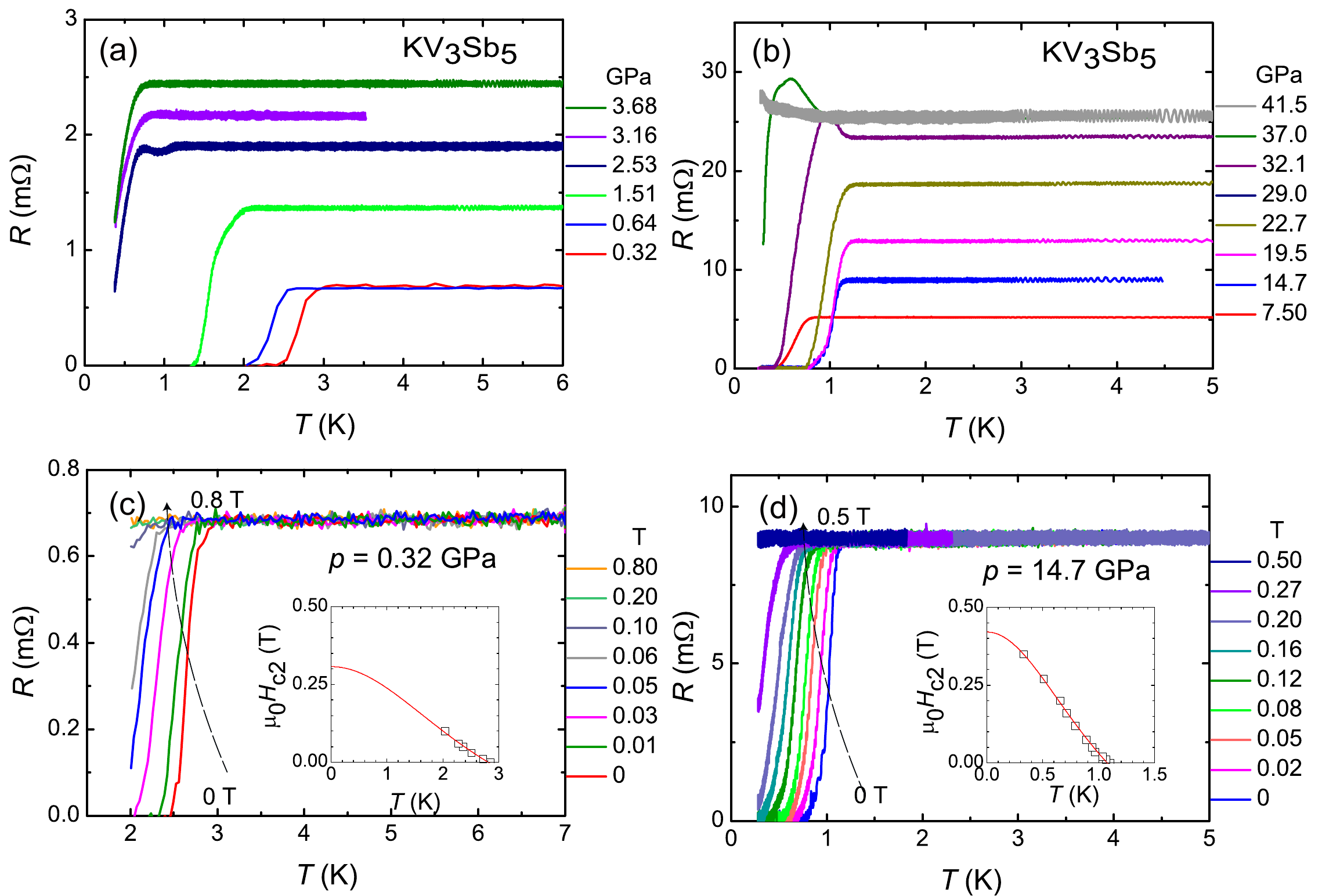}
\caption{(a) Temperature dependence of resistance for KV$_3$Sb$_5$ under various pressures up to 3.68 GPa. (b) Temperature dependence of resistance for KV$_3$Sb$_5$ under higher pressures up to 41.5 GPa. (c) and (d) Temperature dependence of resistance for KV$_3$Sb$_5$ under different magnetic fields at 0.32 GPa and 14.7 GPa, respectively. Inset: Temperature dependence of the upper field $\mu$$_0$${\it H}_{\rm c2}$.}
\end{figure}

\begin{figure}
\includegraphics[clip,width=6cm]{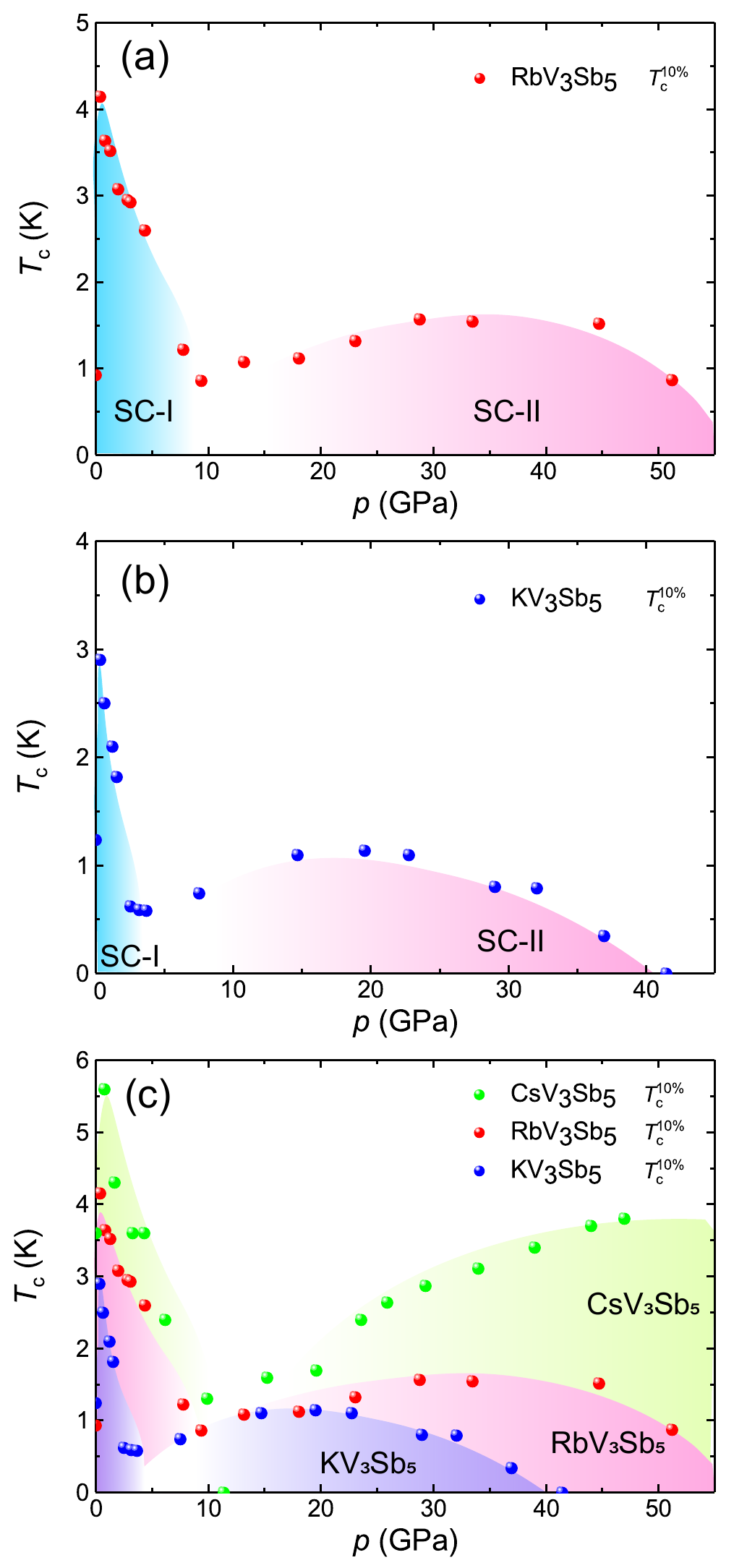}
\caption{ Temperature\textendash pressure phase diagram of (a) RbV$_3$Sb$_5$ and (b) KV$_3$Sb$_5$. Two superconducting domes can be clearly seen for both compounds, with SC-I phase under lower pressures and SC-II phase under higher pressures. (c) Temperature\textendash pressure phase diagram of $A$V$_3$Sb$_5$ ($A$ = Rb and K). The data of CsV$_3$Sb$_5$ are taken from Ref. \cite{SYLiCsVSb}. The phase diagrams of CsV$_3$Sb$_5$, RbV$_3$Sb$_5$ and KV$_3$Sb$_5$ are colored with green, pink, and blue, respectively.}
\end{figure}

Based on above high-pressure resistance measurements of RbV$_3$Sb$_5$ and KV$_3$Sb$_5$, the temperature-pressure ($T$-$p$) phase diagrams are constructed in Figs. 4 (a) and 4(b). Two superconducting domes can be clearly seen for both compounds, named as SC-I under lower pressures and SC-II under higher pressures. Therefore, all three compounds of $A$V$_3$Sb$_5$ ($A$ = Cs, Rb and K) have two superconducting domes under pressure, as a common feature. An integrated ${\it T}_{\rm c}$\textendash $p$ phase diagram of V-based superconductors $A$V$_3$Sb$_5$ ($A$ = Cs, Rb, and K) is summarized in Fig. 4(c). The high-pressure resistance data of CsV$_3$Sb$_5$ are taken from ref. \cite{SYLiCsVSb}. One can see that CsV$_3$Sb$_5$ possesses the largest span phase diagram while KV$_3$Sb$_5$ possesses the smallest one among these V-based superconductors. With increasing the ionic radius of intercalated metal atoms from K to Cs, the maximum ${\it T}_{\rm c}$ in both SC-I and SC-II phases increases. The superconducting transition of CsV$_3$Sb$_5$ is also robust against pressure to more than 50 GPa in the second dome, whereas the superconducting transition of KV$_3$Sb$_5$ is undetectable at 41.5 GPa. Further experimental and theoretical works are needed to clarify the effect of different intercalated alkali-metal atoms on the superconductivity in $A$V$_3$Sb$_5$.

Previously, the pressure-induced double-dome superconductivity was also observed in some other unconventional superconductors. The triggers of the second superconducting dome are categorized as: (i) first-order volume collapse or structural phase transition like in the heavy-fermion superconductor, CeCu$_2$Si$_2$ \cite{CeCu2Si2} and (ii) Lifshitz transition in the Fermi surface such as in iron chalcogenides \cite{L.L.Sun,J.-G. Cheng-LiFeSe} and cuprate superconductors \cite{cuprate}. In V-based superconductors CsV$_3$Sb$_5$ and KCsV$_3$Sb$_5$, the origin of the SC-I phase is strongly related to the CDW instability \cite{SYLiCsVSb,J.-G. Cheng,Pressure-KVSb}. Since there is no any structural phase transition under pressure for CsV$_3$Sb$_5$, the SC-II phase was attributed to a pressure-induced Lifshitz transition and enhanced electron-phonon coupling in CsV$_3$Sb$_5$ \cite{Z. Yang,X. Chen}. The underlying superconducting pairing mechanism of these two superconducting domes needs further investigation.

In summary, we clarify the pressure dependence of superconductivity in the V-based superconductors $A$V$_3$Sb$_5$ ($A$ = Rb, and K). Two superconducting domes are unambiguously revealed in both compounds, as in CsV$_3$Sb$_5$, suggesting the pressure-induce double-dome superconductivity is a common feature in these V-based superconductors. The phase diagram is strongly affected by the ionic radius of intercalated alkali-metal atoms. We hope our results in this work will shed a light on systemical understanding the superconducting pairing mechanism of V-based topological Kagome metals.\\

This work was supported by the Natural Science Foundation of China (Grant No. 12034004), the Ministry of Science and Technology of China (Grant No.: 2016YFA0300503), and the Shanghai Municipal Science and Technology Major Project (Grant No. 2019SHZDZX01). Y. F. Guo was supported by the Major Research Plan of the National Natural Science Foundation of China (No. 92065201) and the Program for Professor of Special Appointment (Shanghai Eastern Scholar). H. C. Lei was supported by National Natural Science Foundation of China (Grant No. 11822412 and 11774423), the Ministry of Science and Technology of China (Grant No. 2018YFE0202600 and 2016YFA0300504), and Beijing Natural Science Foundation (Grant No. Z200005).\\

C. C. Zhu, X. F. Yang, W. Xia, and Q. W. Yin contributed equally to this work.

\noindent $^\dag$ E-mail: hlei$@$ruc.edu.cn\\
\noindent $^\ddag$ E-mail: guoyf$@$shanghaitech.edu.cn\\
\noindent $^\sharp$ E-mail: yangxiaofan@fudan.edu.cn\\
\noindent $^*$ E-mail: shiyan$\_$li$@$fudan.edu.cn

\end{document}